\documentclass[preprint,nofootinbib,floatfix,aps,prd,showpacs,amsmath,amssymb,amsfonts,superscriptaddress]{revtex4-1}
\usepackage{graphicx}
\usepackage{color}
\usepackage{dsfont}
\usepackage[normalem]{ulem}
\usepackage{hyperref}
\usepackage{mathtools}
\usepackage{mathrsfs}
\usepackage{calrsfs}
\usepackage{comment}
\usepackage[utf8]{inputenc}
\usepackage{multirow} 
\usepackage{subfigure}
\usepackage{varioref}
\usepackage[active]{srcltx}

\expandafter\ifx\csname package@font\endcsname\relax\else
\expandafter\expandafter
\expandafter\usepackage
\expandafter\expandafter
\expandafter{\csname package@font\endcsname}%
\fi
\hypersetup{
	colorlinks=true,       
	linkcolor=blue,          
	citecolor=blue,        
}

\usepackage[section]{placeins}

\usepackage{fancybox}
\labelformat{figure}{Fig.~#1}

\begin{document}
\pagenumbering{gobble} 
\title{
Fast Radio Bursts {\it signal} high-frequency gravitational waves
}
\author{Ashu Kushwaha} 
\email{ashu712@iitb.ac.in}
\affiliation{Department of Physics, Indian Institute of Technology Bombay, Mumbai 400076, India}
\author{Sunil Malik}
\email{sunil.malik@uni-potsdam.de}
\affiliation{Department of Physics, Indian Institute of Technology Bombay, Mumbai 400076, India}
\affiliation{Institut f\"ur Physik und Astronomie, Universit\"at Potsdam, Potsdam, Germany}
\affiliation{Deutsches Elektronen-Synchrotron DESY, Zeuthen, Germany}
\author{S. Shankaranarayanan
}
\email{shanki@iitb.ac.in}
\affiliation{Department of Physics, Indian Institute of Technology Bombay, Mumbai 400076, India}
\begin{abstract}
There is growing evidence for high-frequency gravitational waves (HFGWs) ranging from MHz to GHz. Several HFGW detectors have been operating for over a decade, and two GHz events have been reported recently. However, a confirmed detection might take a decade. This essay argues that unexplained observed astrophysical phenomena, like Fast Radio Bursts (FRBs), might provide indirect evidence for HFGWs. In particular, using the Gertsenshtein-Zel$'$dovich effect, we show that our model can explain three key features of FRBs: generate peak-flux up to $1000~{\rm Jy}$, naturally explain the pulse width and the coherent nature of FRBs. In short, our model offers a novel perspective on the indirection detection of HFGWs beyond current detection capabilities. Thus, transient events like FRBs are a rich source for multi-messenger astronomy. 

\vspace*{2.0cm}
\begin{center}
{\bf Essay received honorable mention in Gravity Research Foundation 2023. \\ Matches the published version in \href{https://doi.org/10.1142/S0218271823420105}{IJMPD}.}
\end{center}
\end{abstract}
\maketitle

\newpage
\pagenumbering{arabic} 

It took a century to directly detect gravitational waves (GWs) in the audio frequency range~\cite{2016-LIGOScientific-PRL}. However, in 1974, Hulse and Taylor indirectly detected GWs using electromagnetic (EM) waves in the radio frequency through the binary pulsar PSR B1913+16~\cite{1975-Hulse.Taylor-ApJL}. The indirect detection used the age-old principle --- \emph{energy conservation}. It was discovered that the trajectory of a pulsar around a neutron star (NS) gradually contracts with time. This decay in the orbital period matches the loss of energy and momentum in the gravitational radiation predicted by general relativity and the energy released in the form of GWs.

GWs can be generated in various frequency ranges~\cite{2000-Maggiore-PRep,Bisnovatyi-Kogan:2004cdg,2006-Cruise.Ingley-CQG,2009-Sathyaprakash.Schutz-LivRevRel,2012-Cruise-CQG,2016-Dong.Kinney.Stojkovic-JCAP,2019-Christensen-RPP,2021-Aggarwal.etal-LivingRevRel,2021-Domcke.Garcia-Cely-PRL}. In general, the characteristic frequency of the GW from a compact object is inversely related to object radius $R$ and directly related to mass $M$~\cite{2009-Sathyaprakash.Schutz-LivRevRel}: 
\begin{align}\label{eq:f_0}
    f_0 \propto \frac{1}{4\pi} \left( \frac{3 M}{R^3} \right)^{1/2}
\end{align}
In other words, the smaller the object size, the larger the frequency of the GW~\cite{2000-Maggiore-PRep,Bisnovatyi-Kogan:2004cdg,2006-Cruise.Ingley-CQG,2009-Sathyaprakash.Schutz-LivRevRel,2012-Cruise-CQG,2016-Dong.Kinney.Stojkovic-JCAP,2019-Christensen-RPP,2021-Aggarwal.etal-LivingRevRel,2021-Domcke.Garcia-Cely-PRL}. This is similar to the EM radiation where the frequency depends on the characteristic size of the system that generates it~\cite{Book-Jackson-Classical_Electrodynamics}. However, at the beginning of the 20th century, we could only observe the Universe in a narrow band through EM waves. 
With the advent of radio waves, we now have a wide band of frequencies spanning twenty-orders of magnitude from radio to gamma rays. 
In that sense, GW astrophysics is in the same stage as astronomy at the start of the 1900s.
Similar to how the different spectral ranges of the EM spectrum allow us to understand the Universe differently and to comprehend various events, GWs with \emph{disparate frequencies} can provide distinct perspectives on the cosmos. 

As mentioned above, several physical systems can generate GWs in a broad range of frequencies, $10^{-18} - 10^{10}~{\rm Hz}$~\cite{2000-Maggiore-PRep,Bisnovatyi-Kogan:2004cdg,2006-Cruise.Ingley-CQG,2009-Sathyaprakash.Schutz-LivRevRel,2012-Cruise-CQG,2016-Dong.Kinney.Stojkovic-JCAP,2019-Christensen-RPP,2021-Aggarwal.etal-LivingRevRel,2021-Domcke.Garcia-Cely-PRL} and direct measurements from different experiments can probe a variety of sources.
Since the GW frequency is related to the mass and radius of the object differently~\cite{2009-Sathyaprakash.Schutz-LivRevRel}, and GW weakly couples to matter, one can probe the strong and weak gravity regions to unprecedented accuracy~\cite{2000-Maggiore-PRep,Bisnovatyi-Kogan:2004cdg,2006-Cruise.Ingley-CQG,2009-Sathyaprakash.Schutz-LivRevRel,2012-Cruise-CQG,2016-Dong.Kinney.Stojkovic-JCAP,2019-Christensen-RPP,2021-Aggarwal.etal-LivingRevRel,2021-Domcke.Garcia-Cely-PRL}.  
LIGO-VIRGO-KAGRA operating at frequencies less than a few kHz have detected around 100 GW events from merging black-holes and NSs~\cite{2019-GWTC1-PRX,2021-GWTC2-PRX,2021-GWTC3-Arxiv}. Similarly, nanoGrav and space missions (like LISA and DECIGO) are designed to detect GWs in nanoHz and mHz frequencies~\cite{NANOGrav:2020gpb,2009-Sathyaprakash.Schutz-LivRevRel}. 
Primordial black-holes, Exotic compact objects, and the early Universe physics generate high-frequency GWs (HFGWs) in MHz to GHz range~\cite{2008-Akutsu.etal-PRL,2008-Nishizawa.etal-PRD,2016-Holometer-PRL,2017-Chou.etal-PRD,2020-Ito.etal-EPJC,2021-Domenech-EPJC,2021-Goryachev.etal-PRL,2022-Aggarwal.etal-PRL}. Over the last decade, many HFGW detectors have been proposed, some of which are operational. For instance, the Japanese 100 MHz detector with a $0.75~{\rm m}$ arm-length interferometer has been operational for a decade~\cite{2008-Akutsu.etal-PRL,2008-Nishizawa.etal-PRD}, Holometer detector has put some limit on GWs at MHz~\cite{2016-Holometer-PRL,2017-Chou.etal-PRD}. The Bulk Acoustic GW detector experiment recently reported two MHz events after 153 days of operation~\cite{2021-Goryachev.etal-PRL}. A GHz GW detector is also proposed~\cite{2020-Ito.etal-EPJC}. 

These detectors are ideally suited for searching for physics beyond the standard model, like primordial black-holes, exotic compact objects, and the early Universe~\cite{2021-Aggarwal.etal-LivingRevRel}. However, these detectors are in the early stages, and it may take a decade to make a confirmed HFGW detection. This leads us to the following questions: If certain astrophysical mechanisms generate HFGWs,  how can they be detected? If we cannot detect these waves directly, do astrophysical phenomena indicate their existence? 
The kHz GWs events are very short and carry a lot of energy. For instance, during the final moments of 
LIGO’s first detection lasting 0.2 seconds, more energy ($10^{20} \rm{Jansky}$) was radiated than the light from all the stars in the galaxy~\cite{2016-LIGOScientific-PRL,2009-Sathyaprakash.Schutz-LivRevRel}. Extrapolating these features, we see that to detect HFGW signals, one has to look for highly energetic astrophysical events that last for less than a second.

As the binary-neutron star system was a smoking gun for audio-frequency GWs, are there any transient and highly energetic astrophysical phenomena of unknown origin yet?
In the rest of this essay, we argue that Fast radio bursts (FRBs) --- extremely energetic millisecond-burst of radio emission produced by the galactic and extra-galactic sources~\cite{2008-Lorimer-LivRevRel,2019-Cordes.Chatterjee-AnnRevAA,2019-Platts.etal-PhyRept} 
--- can be the \emph{smoking gun} for the existence of HFGWs. 

To date, around 700 FRBs have been reported in various catalogs in the wavelength range of 800 - 1400 MHz~\cite{Petroff:2016tcr,2019-Platts.etal-PhyRept, Pastor-Marazuela:2020tii,2021-Rafiei.etal-APJ,2021-CHIME-FRB-arXiv}. $99\%$ of these FRBs have the following three characteristic features: (i) observed peak flux ($S_{\nu}$) in the range $0.1~{\rm Jy} < S_{\nu} < 700~{\rm Jy}$, (ii) coherent radiation and (iii) pulse width that is less than a second \cite{2021-Rafiei.etal-APJ,Petroff:2016tcr}. Since the time scale of these events is less than a second, and the emission is coherent, the astrophysical mechanisms that explain these events \emph{cannot} be thermal~\cite{2008-Lorimer-LivRevRel}. Many models involving non-thermal processes such as Synchrotron radiation~\cite{Book-Lorimer.Kramer-PulsarAstronomy}, black hole super-radiance~\cite{2018-Conlon.Herdeiro-PLB}, evaporating primordial black hole \cite{1977-Rees-Nature,2020-Carr.Kuhnel}, spark from cosmic strings~\cite{2008-Vachaspati-PRL}, Quark Novae~\cite{2015-Shand.etal-RAA},  {synchrotron maser shock model~\cite{2020ApJ...900L..26W},  radiation from reconnecting current
sheets in the far magnetosphere~\cite{2020ApJ...897....1L},  curvature emission from charge bunches~\cite{2022ApJ...927..105W} } have been proposed. However, despite the use of exotic new physics, no single model has been able to provide a universal explanation for the enormous energy released in these events. 

Due to the nature of electromagnetic interaction, small-scale emission mechanisms usually predominate over large-scale coherent electromagnetic processes (like astrophysical masers and pulsar radio emission). Hence, a mechanism that explains FRBs needs to clarify how to generate a large amount of coherent radiation in a short 
time~\cite{2018-Popov.etal-Usp,2022-Lieu.etal-CQG}. This is where gravity helps! While GWs are also generated by accelerating masses, there is one fundamental difference between GWs and EM waves. All masses have the same gravitational sign and tend to clump together to produce large coherent bulk motions that generate \emph{energetic, coherent GWs}~\cite{2007-Hendry.Woan-AG}. Thus, if a mechanism that converts incoming coherent GWs to EM waves exists, we can explain the extremely energetic, coherent nature of FRBs. 

Given that GWs as a source can explain coherence, we ask the following questions: Is there a physical mechanism that converts the incoming coherent GWs to EM waves? If that exists, can the necessary conditions for such a mechanism naturally be found in any astrophysical object/region? GWs get converted into EM waves in the presence of a strong transverse magnetic field via the Gertsenshtein-Zel$'$dovich (GZ) effect~\cite{1962-Gertsenshtein-JETP,1974-Zeldovich-SJETP,2021-Domcke.Garcia-Cely-PRL}. The GWs passing through a region with high transverse magnetic field ${\bf B}$ leads to compression and stretching of the magnetic field proportional to $h {\bf B}$ ($h$ is the amplitude of GWs), which acts as a source leading to the generation of EM waves. The resultant EM waves generated will have maximum amplitude at resonance (same frequency as incoming GWs)~\cite{1962-Gertsenshtein-JETP,1974-Zeldovich-SJETP,2021-Domcke.Garcia-Cely-PRL}. In quantum mechanical language, the GZ effect is analogous to the mixing of neutrino flavors --- the external field \emph{catalyzes} a resonant mixture of photon and graviton states~\cite{2023-Palessandro.Rothman-PDU}. The external magnetic field provides the extra angular momentum necessary for the spin-1 (photon) field to mix with the spin-2 (graviton) field. 

\begin{figure*}[!ht]
\centering
\includegraphics[height=2in]{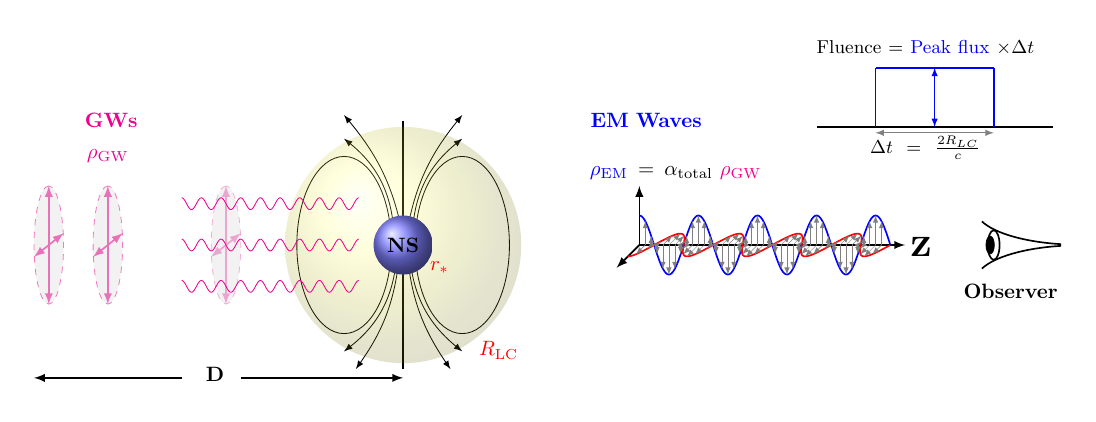}%
\caption{Schematic depiction of the model to explain FRBs due to HFGWs. The blue sphere is NS with radius $r_*$ and the yellow region around NS corresponds to magnetosphere of radius $R_{\rm LC}$. The black curves correspond to the magnetic field lines. At resonance, the incoming HFGWs (magenta) transmutes to EM waves (blue) in the magnetosphere.}
\label{fig:setupFigure}
\end{figure*}

Interestingly, NSs are compact objects with large magnetic fields ($> 10^{8}\ {\rm Gauss}$), which can lead to a significant conversion of GWs to EM waves. As shown in \ref{fig:setupFigure}, consider \emph{transient, coherent GWs} produced by exotic compact objects (such as Boson stars, Oscillons, gravastars) passing through the magnetosphere of NS at a distance $D$ (where $D > 10~{\rm kpc}$). The cylinder around the NS encloses the magnetosphere. GZ effect converts GWs to EM waves as they pass through the magnetosphere~\cite{1962-Gertsenshtein-JETP,1974-Zeldovich-SJETP}. This conversion occurs at all points in the magnetosphere. Consequently, a distant observer (in the z-axis) will perceive this effect occurring in the entire magnetosphere during this brief period. Since the conversion happens at the resonance frequency, as shown below, the FRBs' energy flux depends on the amplitude of the incoming GWs, the magnetic field's strength, and light cylinder radius ($R_{\rm LC})$. 
[$R_{\rm LC}$ denotes the location where the co-rotation velocity equals the speed of light (see \ref{fig:setupFigure}).]
Thus, our FRB model has three key ingredients: (i) incoming plane HFGWs, (ii) magnetosphere around the NS, and (ii) conversion from GWs to EM waves. Before we evaluate the energy flux and compare it with observations, we discuss the assumptions about these three key ingredients. 

As mentioned earlier, since accelerating masses clump together, they produce large coherent bulk motions that generate \emph{energetic, coherent GWs}~\cite{2007-Hendry.Woan-AG}. Since we assume NS, to be at a distance $D$, in the radiation zone of the GW source, the GWs can be treated as plane waves~\cite{2022-Lieu.etal-CQG}. Thus, the line element of a $+$ (and similarly for $\times$) polarization mode of GW propagating (with frequency $\omega_g$ and wave-vector $k_g$) along the z-direction, is:
\begin{align}\label{eq:metric}
    ds^2 =  d(ct)^2 - (1+h_+) dx^2 - (1-h_+)dy^2 - dz^2
\end{align}
where,
\begin{align}\label{eq:h-Expression}
h_+  = 
A_+ \, e^{i \left( k_g z - \omega_g t \right) }, 
h_{\times}  = 
i A_{\times} \, e^{i \left( k_g z - \omega_g t \right)} 
,
\end{align}
$A_+$ and $A_{\times}$ are the constant GW amplitudes. In the linear order, the two polarization modes evolve independently and have equal magnitudes~\cite{Chandrasekhar_BlackHoles-Book}. The characteristic strain ($h$) for exotic compact objects, with compactness $C$, is~\cite{2021-Aggarwal.etal-LivingRevRel}:
\begin{equation}
h \lesssim  10^{-19} C^{5/2} \left( \frac{\rm MHz}{\omega_g} \right) \left( \frac{\rm Mpc}{D} \right) \,  .
\end{equation}
This translates to  
 \[
 h_{\rm 1.4 GHz, 10kpc} \lesssim 10^{-21} \, ,
  h_{\rm 1.4 GHz, 1Mpc} \lesssim 10^{-23} \ .
 \]
In our model, we assume the GW amplitude $h$ at the NS magnetosphere (at $10~{\rm kpc}$ distance) to be $10^{-23}$ which is two orders of magnitude \emph{lower than} the above-estimated amplitude. 

NS generally has a dipolar magnetic field and is likely surrounded by a magnetosphere free from other forces~\cite{Book-Lorimer.Kramer-PulsarAstronomy}. At any point in the magnetosphere, we approximate the magnetic field by a constant value~\cite{2022-Kushwaha.etal-arXiv}. The effective time-dependent transverse magnetic field at a given point in the NS magnetosphere is given by
\begin{align}
\textbf{B}(t) = \left( 0, B^{(0)}_y + \delta B_y \sin (\omega_B t), 0  \right)
\end{align}
where $\omega_B$ is the frequency alternating magnetic field, which is equal to the frequency of rotation of NS (with $\omega_B \ll \omega_g$)~\cite{2012-Pons.etal-AandA}. The small time-dependence arises due to the rotation of the NS about its axis with frequency $\omega_B$~\cite{2019-Platts.etal-PhyRept,2019-Pons.Vigan-arXiv,2012-Pons.etal-AandA}. We consider $\omega_B$ in the range $[1, 10^3]~{\rm Hz}$~\cite{Book-Lorimer.Kramer-PulsarAstronomy}\footnote{In this work, we use frequency and angular frequency interchangeably, hence we have used Hz.}. 
Note that $\xi \equiv \delta B_y/B_y^{(0)}$  can be as large as $0.1$~\cite{2012-Pons.etal-AandA}. We assume $\xi \sim  10^{-2}$.

To evaluate the transmutation of GWs to EM waves, we need to solve the covariant Maxwell's equations in the background metric \eqref{eq:metric}. At linear-order, the  induced electric and magnetic fields are~\cite{2022-Kushwaha.etal-arXiv}: 
\begin{align}\label{eq:E_x-Final}
\tilde{E}_x &\simeq  - \frac{A_{+} }{2} \, B^{(0)}_y \left( 1   -  \xi \,  \omega_B t  \,\, \right) \,\, e^{i \left(k_g z - \omega_g t\right) } 
\\
\label{eq:B_y-Final}
\tilde{B}_y &\simeq  - \frac{A_{+} }{4} \, B^{(0)}_y \left(1  +  2 \xi \,  \omega_g t  \,\,\right) e^{i \left(k_g z - \omega_g t\right) } \, .
\end{align}
Since the incoming GWs are plane-polarized, the transmuted EM waves will also be plane-polarized and, will have the same frequency at resonance~\cite{1962-Gertsenshtein-JETP,2023-Palessandro.Rothman-PDU}.

A conversion factor ($\alpha_{\rm{tot}}$), or ratio of the energy density of EM waves to GWs, can be used to estimate the conversion efficiency. For the entire magnetosphere, we get~\cite{2022-Kushwaha.etal-arXiv}:
\begin{align}\label{eq:alpha_tot}
\alpha_{\rm{tot} } \simeq
\frac{5 \pi G |B^{(0)}_y|^2 }{2 c^2} \left[  \frac{4}{15} \xi^2  \left[\frac{ R_{\rm LC}}{c} \right]^2 \!\! + \frac{2 \xi R_{\rm LC}}{5 \omega_g c} + \frac{1}{\omega_g^2} \right].
\end{align}

Having discussed the assumptions of the model, in the rest of this essay, we discuss the key predictions of the model and compare them with the observations. Spectral flux density is the key observed quantity in all the FRB catalogs~\cite{Petroff:2016tcr, 2019-Platts.etal-PhyRept,Pastor-Marazuela:2020tii,2021-Rafiei.etal-APJ,2021-CHIME-FRB-arXiv}. This observed quantity is related to the theoretically evaluated Poynting vector per unit frequency~\cite{1979-Rybicki.Lightman-Book}. The Poynting vector is well-defined for photons that travel from the source to the observer without any hindrance~\cite{1979-Rybicki.Lightman-Book,Book-Carroll.Ostlie-CUP,Book-Condon.Ransom-PUP,Book-Zhang-GRB-CUP}. The Poynting vector estimated at a small angle (along the direction of the incoming gravitational waves) remains the same at the source and the detector. Thus, the energy flux carried by the induced EM waves is~\cite{2022-Kushwaha.etal-arXiv}:
\begin{align}\label{eq:poyntingVector}
    S_z = \frac{c}{8 \pi} \tilde{E}_x \times \tilde{B}^*_y \simeq \frac{A_+^2 \, |B_y^{(0)}|^2  \, c }{64 \pi} \left[ 1 + 2 \omega_g \xi \frac{R_{\rm{LC}}}{c} - 2 \omega_g \omega_B \xi^2 \left( \frac{R_{\rm{LC}}}{c}\right)^2  \right]
\end{align}
where $\tilde{B}^*_y$ is the complex conjugate of the induced magnetic field $\tilde{B}_y$. 

As mentioned above, the Poynting vector per unit frequency is the spectral flux density that characterizes FRBs~\cite{Petroff:2016tcr,2019-Platts.etal-PhyRept, Pastor-Marazuela:2020tii,2021-Rafiei.etal-APJ}. The table below contains numerical values predicted by our model for the three input parameters --- $\omega_g$ (frequency of the incoming HFGW), $R_{\rm LC}$ (light cylinder radius that defines the magnetosphere), and $B_y^{(0)}$ (average magnetic field in the magnetosphere). We see that our model predicts a range of spectral flux density that can be as small as $0.1~{\rm Jy}$ (for milli-second Pulsar) and can be as large as $10^{11}~{\rm Jy}$ (for Magnetar). 

\begin{table}[h]

\centering
\begin{tabular}{|c|c|c|c|c|c|c|}
	\hline
	    $R_{\rm LC }$   &  $B_y^{(0)}$    &   $\omega_g $    &  $\alpha_{\rm tot}$  & $\rho_{\rm EM} \,  $ & $\frac{S_z}{\omega_g}$ & Fluence \\ 
	   (cm)  &  (Gauss)  &  (MHz) &  &  ($\rm{Jy \, cm^{-1} \, s \, Hz }$)  &  (Jy)  & (Jy ms)\\
	\hline
   $10^9$ &  $10^{15}$  & $1$  & $1.74 
   \times 10^{-5}$ & $4.65 \times 10^{10} $ & $9.95\times 10^{11}$ & $6.62 \times 10^{13}$\\ 
 \hline
   $10^9$  &  $10^{12}$  & $500$  &  $ 1.72 \times 10^{-11}$  & $1.15 \times 10^{10}  $ & $9.94\times 10^{5}$ & $6.62 \times 10^{7}$\\ 
 \hline
 $10^9$  &  $5 \times 10^{11}$  & $1250$  &  $ 4.3 \times 10^{-12}$  & $1.8 \times 10^{10}  $ & $2.48\times 10^{5}$ & $1.65 \times 10^{7}$\\ 
 \hline
     $10^8$  &  $10^{11}$  & $1400$  &  $ 1.72 \times 10^{-15}$  & $9.07 \times 10^6  $ & $ 961.57 $ & $6410.46$ \\ 
 \hline
   $10^7$  &  $10^{10}$  & $1400$  &  $1.72 \times 10^{-19}$  & $9.07 \times 10^{2} $ & $0.99$  & $0.66$\\ 
 \hline
   $10^8$  &  $10^{9}$  & $1400$  &  $1.72 \times 10^{-19}$ &  $9.07 \times 10^{2} $ & $0.09$ & $0.6$ \\ 
 \hline
 %
      %
  ${10^9}^{*}$  &  $2.2\times 10^{14}$  & $1400$  &  $8.34 \times 10^{-7}$ &  $4.39 \times 10^{9} $ & $4.8\times 10^4$ & $3.2\times 10^6$ \\ 
 \hline
\end{tabular}
\caption{Numerical values of the total conversion factor ($\alpha_{\rm tot}$), energy density of EM waves ($\rho_{\rm EM}$), and spectral flux density (Poynting vector per unit frequency). The first two rows are for a typical Magnetar and the last three rows are for a typical NS. Last column shows the observed fluence which is the product of the burst width $\Delta t = (2R_{LC})/c$ and the peak flux ($S_z/\omega_g$)~\cite{2015-Keane.Petroff-MNRAS,2018-Macquart.Ekers-MNRAS}. 
We have set  $G = 6.67 \times 10^{-8}~{\rm dyne \, cm^2 gm^{-2}},
c = 3 \times 10^{10} {\rm cm \,\, s^{-1}}, \, A_{+} = 10^{-23}$ corresponding to a typical GW source. This row represents the fluence estimated corresponds to the FRB 200428 using the $A_+ = 10^{-26}$, which is associated with the galactic magnetar SGR 1935+2154~\cite{Bochenek2020Natur,2020Natur.587...54C}. }
\label{table1}
\end{table}

This is the most important conclusion of this essay, regarding which we wish to emphasize the following points: 
\begin{enumerate}
\item From the penultimate column of Table \eqref{table1}, we see that our model predicts the burst of EM wave with the flux $< 1000~\rm{Jy}$. Our model predicts that the progenitor should be a NS with an effective magnetic field strength in the range $10^{9} - 10^{11}~{\rm G}$ and rotation frequency $1 < \omega_B < 1000~\textcolor{blue}{\rm{Hz}}$. Thus, our model has the potential to explain the observed peak flux of most of the reported non-repeater FRBs~\cite{Petroff:2016tcr, 2019-Platts.etal-PhyRept,Pastor-Marazuela:2020tii,2021-Rafiei.etal-APJ,2021-CHIME-FRB-arXiv}.

\item In particular, in recent observation, a magnetar SGR 1935+2154 residing in Milky Way is reported to be associated with the FRB 200428 with fluence $>1.5 \times 10^6$ Jy ms by the STARE2 radio array
in $1.28–1.4~$GHz band~\cite{Bochenek2020Natur,2020Natur.587...54C}. This magnetar is reported to have a surface dipole magnetic field of $\sim2.2\times10^{14}$ G, which is inferred from the slow-down rate of period $3.24 \ {\rm s}$. If we consider $B_y^{(0)} \sim 5 \times 10^{11}$ G as a static approximation with ${\rm R_{LC}} \sim 10^9$ cm, our model can predict the fluence of the order of $1.65 \times 10^7 \rm{Jy \, ms}$. Thus it can potentially explain these observations and play a significant role in future FRB and magnetar association (if any).

\item Recently, in Ref. \cite{Kalita:2022uyu}, the authors used the GZ effect to compare the predicted output of the GZ signal with two known FRBs and their possible detection in GW detectors. They showed that if the proposed GW detectors detect any continuous GW signal from the site of FRBs, this will immediately imply that the merger-like theories cannot explain all FRBs and thus provide significant support for the GZ theory.


\item It is estimated that around $10^8 - 10^9$ NSs are present in any given galaxy like the Milky Way, roughly $1\%$ of the total number of stars in the galaxy~\cite{2006-Diehl.etal-Nature,2010-Sartore.etal-AandA}. Also, it is estimated that the magnetar formation rate is approximately $1 - 10$ percent of all pulsars~\cite{2015-Gullon.etal-MNRAS,2019-Beniamini.etal-MNRAS}. Since Magnetars are rare, our model predicts that observing such a large spectral flux density is unlikely.

\item One of the fundamental assumptions of our model is that, given a GW signal in the GHz frequency range, all NSs always act as FRB sources.
This assumption means the maximum FRB events per day can be $10^8 - 10^9$. However, the estimated FRB rate is $10^{3}$ for the entire sky per day~\cite{Champion2016MNRAS}. This can be attributed to the fact that the probability of this event is a product of the probability that the GW passes through the NS times the probability that the emitted EM is along the line of sight of the observer.

\item $R_{\rm LC}$ for a typical NS is $\sim 10^{7}-10^{9}~{\rm cm}$, implying that the GWs take less than one second to cover the entire magnetosphere~\cite{Book-Lorimer.Kramer-PulsarAstronomy}. This implies that the induced EM waves due to the GZ effect will appear as a burst lasting for less than one second. Thus, our model provides a natural explanation for the pulse width (lasting less than a second) of FRBs.

\item Since the incoming GWs are coherent, the induced EM waves will also be coherent. Thus, our model also explains the coherent nature of FRBs.
\end{enumerate}

In this essay, our main focus is to propose and demonstrate that the FRB can act as an indirect detector for high-frequency gravitation waves, which are not repeaters. If multiple GWs sources are located near the magnetosphere of the neutron stars, in principle, the GZ mechanism should explain the repeaters' FRB and their aperiodicity. However, this needs further investigation.

There is growing evidence for the existence of HFGWs ranging from MHz to GHz. Our results indicate that FRBs can be the smoking gun for the existence of HFGWs. As we just showed, in this model, the observed FRBs are due to the transmutation of GWaves to EM waves in the GHz range. Since the GZ mechanism has no complex physics, FRBs can provide key information about the sources emitting GWs in the GHz. Thus, a century after General Relativity, GWs with disparate frequencies can provide a distinct perspective of the cosmos, which can help to resolve some of the long-standing problems in cosmology and astrophysics. \emph{Time will tell}. \\[10pt]

\noindent {\bf Acknowledgements}  
We thank the referee for the comments and suggestions that greatly improved the manuscript. The authors thank J. P. Johnson for comments on the earlier draft. AK is supported by the MHRD fellowship at IIT Bombay. This work is supported by ISRO Respond grant and SERB-Core Research Grant.

\newpage

\bibliography{Ref-GRF.bib}
	
\end{document}